\documentclass[12pt]{article}
\usepackage{graphicx}

\newcommand\pubnumber{Experiment-1-1}
\newcommand\pubdate{\today}

\def\institute{Department of Physics\\
University of Florida}
\def\authemail{\footnote{Copyright 2023 CERN for the benefit of the ATLAS and CMS Collaborations. Reproduction of this article or parts of it is allowed as specified in the CC-BY-4.0 license}}

\def\Title#1{\begin{center} {\Large #1 } \end{center}}
\def\Author#1{\begin{center}{ \sc #1} \end{center}}
\def\Address#1{\begin{center}{ \it #1} \end{center}}

\newcommand\pubblock{\rightline{\begin{tabular}{l} \pubnumber\\
         \pubdate  \end{tabular}}}
\newenvironment{Abstract}{\begin{quotation}  }{\end{quotation}}
\newenvironment{Presented}{\begin{quotation} \begin{center} 
             PRESENTED AT\end{center}\bigskip 
      \begin{center}\begin{large}}{\end{large}\end{center} \end{quotation}}

\def\beq{\begin{equation}}
\def\eeq#1{\label{#1}\end{equation}}
\def\eeqn{\end{equation}}

\def\beqa{\begin{eqnarray}}
\def\eeqa#1{\label{#1}\end{eqnarray}}
\def\eeqan{\end{eqnarray}}

\let\bar=\overbar

\def\Dslash{\not{\hbox{\kern-4pt $D$}}}
\def\dslash{\not{\hbox{\kern-2pt $\del$}}}

\def\msb{{\bar{\ssstyle M \kern -1pt S}}}

\begin{document}
\begin{titlepage}
\pubblock

\vfill
\Title{Measurements and EFT fits on detector level}
\vfill
\Author{ Kelci Mohrman for the ATLAS and CMS Collaborations\authemail}
\Address{\institute}
\vfill
\begin{Abstract}
The framework of standard model effective field theory (SMEFT) provides a relatively model-agnostic tool with which to parameterize the off-shell effects of potential heavy new physics. In the top quark sector, the ATLAS and CMS experiments make use of data collected by the CERN LHC to search for these off-shell effects by probing dimension-six SMEFT operators involving top quarks. The analyses employ a variety of approaches, which range from relatively indirect (e.g. reinterpretations of unfolded measurements) to relatively direct (in which the SMEFT effects are propagated all the way to detector-level predictions). These proceedings discuss the landscape of detector-level SMEFT analyses in ATLAS and CMS, describe the methodology of the direct detector-level approach, and summarize several recent experimental results.  

\end{Abstract}
\vfill
\begin{Presented}
$16^\mathrm{th}$ International Workshop on Top Quark Physics\\
(Top2023), 24--29 September, 2023
\end{Presented}
\vfill
\end{titlepage}
\def\thefootnote{\fnsymbol{footnote}}
\setcounter{footnote}{0}
%


\section{Introduction}
\label{intro}

The standard model (SM) of particle physics provides a very successful description of the known fundamental particles and their interactions. However, the strong evidence for phenomenon that are not explained by the SM (such as dark matter and dark energy) imply that the SM does not represent a complete description of nature. For this reason, one of the main goals of the CMS~\cite{cmsref,cmsref_r3} and ATLAS~\cite{atlasref} physics programs is to search for signs of new physics beyond the SM. However, at this point, there is not a strong indication of a particular direction in which to search for such evidence. Furthermore, there is no guarantee that new physics particles are light enough to be produced on-shell within the energy range of the CERN LHC, which will not significantly increase throughout its remaining years of operation. The framework of standard model effective field theory (SMEFT) provides a relatively model-agnostic method of parametrizing the off-shell effects of heavy new physics, and can thus provide an interesting method of probing potential new physics at the LHC.

In the SMEFT approach, the SM Lagrangian is treated as the lowest order term in an expansion of higher-order operators that describe new physics at a mass scale referred to as $\Lambda$, where the strength of the new physics interactions are determined by dimensionless parameters referred to as Wilson coefficients (WCs). The SMEFT Lagrangian can thus be written as follows:

\begin{equation}
\label{eq:Left}
    \mathcal{L}_{\mathrm{EFT}} = \mathcal{L}_{\mathrm{SM}} +
    \sum\limits_{d,i} \frac{c^{d}_i}{\Lambda^{d-4}}
    \mathcal{O}^{d}_i. 
\end{equation}

In Equation~\ref{eq:Left}, the $\mathcal{O}^{d}_i$ are the $i$ operators at dimension $d$, the $c^{d}_i$ are the corresponding WCs, and the $\Lambda$ is the mass scale of the new physics. If all of the WCs in Equation \ref{eq:Left} are qual to zero, the SM Lagrangian is recovered, meaning that any non-zero WC would be a sign of new physics. It can also be noted that each term in the expansion is scaled by an additional power of Lambda, so the lowest order terms are expected to contribute most significantly. Here we focus on dimension-six terms, as these are the lowest order terms that contribute to SMEFT in the top quark sector. 

In the SMEFT framework, there are thousands of possible dimension-six operators; the precise number depends on the assumptions that are made (e.g. what flavor symmetries are assumed)~\cite{Henning:2015alf}. Under the flavor universality assumption, the number of dimension-six SMEFT operators is of order 60~\cite{Grzadkowski:2010es}. SMEFT analyses in the top sector often choose to focus on operators involving at least one top quark, with this requirement, the number of operators is reduced to about 40 (though again the precise number depends on the particular assumptions that are made)~\cite{Aguilar-Saavedra:2018ksv}. These operators can be categorized into four main groups based on which particles are involved in the interactions to which they give rise: interactions involving 4 heavy quarks (where heavy refers to top and bottom), interactions involving 2 heavy quarks and 2 light quarks, interactions involving 2 heavy quarks with bosons, and interactions involving 2 heavy quarks and leptons.

\begin{figure}[!h!tbp]
\centering
\includegraphics[height=1.1in]{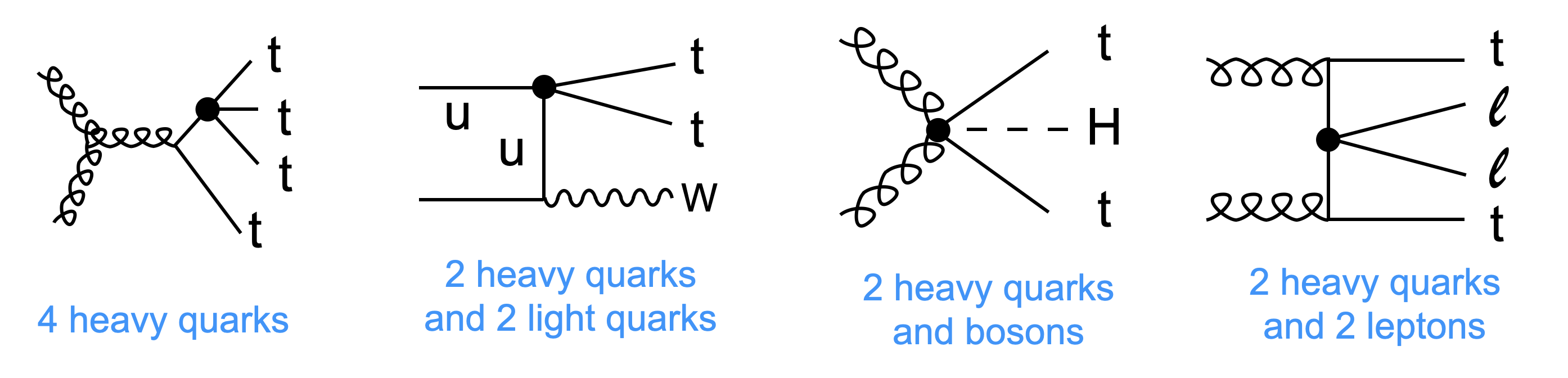}
\caption{Example vertex from each operator category  impacting  an example  process.}
\label{fig:eft_diagrams}
\end{figure}

The operators in each category can impact top quark processes; an example of a vertex from each category is shown in Figure~\ref{fig:eft_diagrams}. The presence of a SMEFT vertex in a top quark processes can impact observables, such as cross sections and kinematic distributions. The strength of these impacts are determined by the value of the WCs. The goal of SMEFT analyses is thus to identify the true value of these WCs, since any non-zero WC would indicate a sign of new physics. In the absence of a signal, the limits placed on the WCs can  be interpreted in terms of the energy scale that is probed, and the limits can also be used to help to constrain theory models. 

\section{Using SMEFT to search for new physics}

In order to find the best fit values for SMEFT WCs, an observable must be parametrized in terms of the WCs in order to obtain a prediction that can be compared against an observation. Let us first consider how the cross section depends on the WCs. To calculate a cross section, one must first write down the matrix element. With the SMEFT framework, this involves not only the SM diagrams, but also the diagrams involving an insertion of a SMEFT vertex (e.g. as shown in Figure~\ref{fig:eft_diagrams}). Squaring the matrix element to obtain the cross section will give rise to three different types of terms: terms in which SM diagrams are multiplied by each other (which have no WC dependence and simply correspond to the SM contribution), terms in which SM diagrams multiply SMEFT diagrams (which contain one power of a WC, and represent the interference between the SM and the SMEFT), and diagrams in which SMEFT diagrams are multiplied by each other (which contain two powers of WCs, and represent the pure new physics contributions). The dependence of the cross section on the WCs is thus represented by an $n$-dimensional quadratic function, where $n$ is the number of WCs. The quadratic dependence will hold for any cross section, inclusive or differential; as long as the SMEFT is modeled linearly in amplitude, a quadratic dependence is expected\footnote{It should be noted that other contributions (e.g. interference between dimension-eight SMEFT operators and the SM, and interference between the SM and diagrams with two insertions of dimension-six SMEFT operators) also enter at the $\Lambda^{-4}$ power. If these $\Lambda^{-4}$ contributions are not able to be included, the quadratic $c^2$ pieces are sometimes dropped from the parameterization (i.e. only the linear pieces are kept) for consistency.}.

The methods which analyses attempt to find the best fit values of WCs generally fall along a spectrum from relatively indirect approaches to relatively direct detector-level approaches. Direct detector-level approaches are the focus of these proceedings. With direct approaches, the SMEFT effects are propagated all the way to detector level, and the WC limits are extracted by comparing the predicted yield in the observable bins against the observed number of events in the bins.  Since yields are proportional to cross sections, the predicted yields also carry a quadratic dependence on the WCs. However, the shape of the quadratic varies from bin to bin (e.g. a bin in the high $p_{\rm T}$ tail of a $p_{\rm T}$ distribution would be expected to have a much steeper quadratic dependence on the WCs than would a bin in the low $p_{\rm T}$ range of the distribution due to the fact that SMEFT effects tend to be larger at higher energies). 

In order to obtain the parameterizations for each bin, detector-level SMEFT analyses often employ the flexible approach of event-by-event reweighting~\cite{Mattelaer:2016gcx}. Each simulated event carries with it a weight, which  corresponds to how much of the cross section that particular event contributes to, so the weight is essentially a small piece of the cross section, or a differential cross section; thus, the weight of each individual event should also carry with it an $n$-dimensional quadratic dependence on the WCs. Technical details describing how this per-event $n$-dimensional quadratic dependence is obtained  and propagated  to detector level are described in reference~\cite{CMS:2020lrr}. Once this per-event quadratic parameterization has been obtained for the detector-level samples, the quadratic dependence for a given bin can  be found by simply summing the quadratics for all of the events that pass the selection criteria for this bin. Since the sum of multiple quadratics is still quadratic,  the predicted yield for an arbitrary observable bin is thus also described by a quadratic in terms of the WCs. 

\begin{figure}[!h!tbp]
\centering
\includegraphics[height=1.9in]{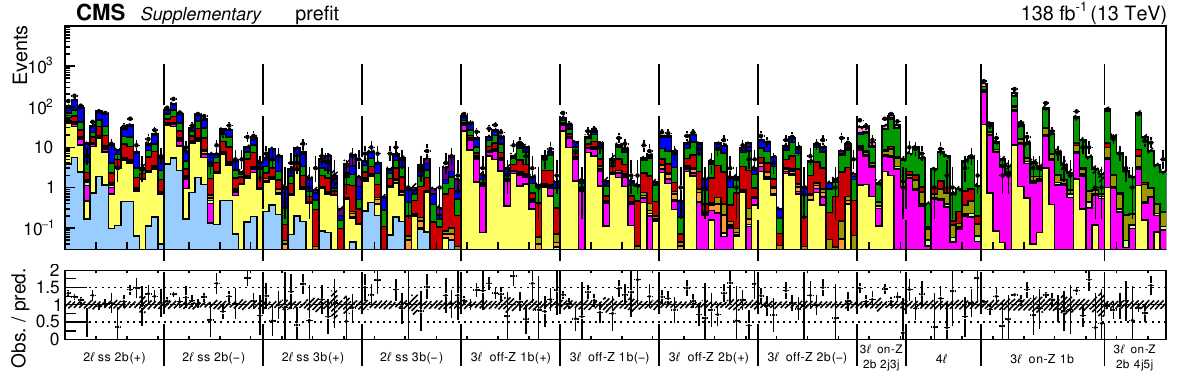}
\caption{Observed data and expected yields in the prefit scenario. ~\cite{CMS:2023xyc}}
\label{fig:mega_plot}
\end{figure}

An example of a detector-level SMEFT analysis which uses the event reweighting approach is the recent CMS multilepton analysis~\cite{CMS:2023xyc}. This analysis studies events involving two same-sign leptons or three or more leptons in order to probe 26 WCs (from all four categories shown in Figure~\ref{fig:eft_diagrams}) simultaneously. Each of the 178 observable bins in the analysis are shown in Figure~\ref{fig:mega_plot}. The predicted yield in each bin is parameterized by a 26-dimensional quadratic function in terms of the 26 WCs. In Figure~\ref{fig:mega_plot}, the histograms have been reweighted to the SM (i.e. the WCs have all been set to zero, and the value in each bin thus corresponds to the constant piece of the quadratic function). This SMEFT-aware histogram can then be passed to the statistical analysis tools, which will adjust the values of the 26 WCs (which will change the shape and normalization across all 178 of the bins) in order to extract the ranges of the 26 WCs which give rise to predicted yields that are in agreement with the observation in each bin. 

While direct-detector level approaches are powerful in that they can provide good sensitivity to SMEFT effects and can offer a consistent and comprehensive modeling of final states involving complicated admixtures of processes and SMEFT effects, there are also challenges associated with the approach. The challenges stem mainly from  the necessity of producing detector-level SMEFT samples. These samples are computationally intensive to produce, and would need to be reproduced if any change to the SMEFT modeling is to be incorporated. For this reason, reinterpretations and analysis preservation may be challenging  for analyses which follow the direct detector-level approach.


\section{Recent detector-level SMEFT analyses in CMS and ATLAS}

The new detector-level SMEFT analyses  (i.e. analyses completed since the Top2022 conference in the preceding year) from ATLAS and CMS are listed below. 
\begin{itemize}
\item CMS: Search for charged lepton flavor violation (involving electrons and muons) in trilepton final states, places limits (individually) on CLFV WCs~\cite{CMS-PAS-TOP-22-005}.
\item ATLAS: Search for charged lepton flavor violation (involving muons and taus) in dilepton final states, places limits (individually) on CLFV WCs~\cite{ATLAS-CONF-2023-001}.
\item CMS: Search for new physics impacting associated top t($\bar{\rm t}$)X processes in multilepton (two same-sign leptons or three or more leptons) final states, places limits (individually and simultaneously) on 26 WCs~\cite{CMS:2023xyc}.
\item ATLAS: Measurement of t-channel production of single top, places a limit on  one WC (from the 2 light  2 heavy  category of operators)~\cite{ATLAS-CONF-2023-026}.
\end{itemize}

\begin{figure}[!h!tbp]
\centering
\includegraphics[height=3.4in]{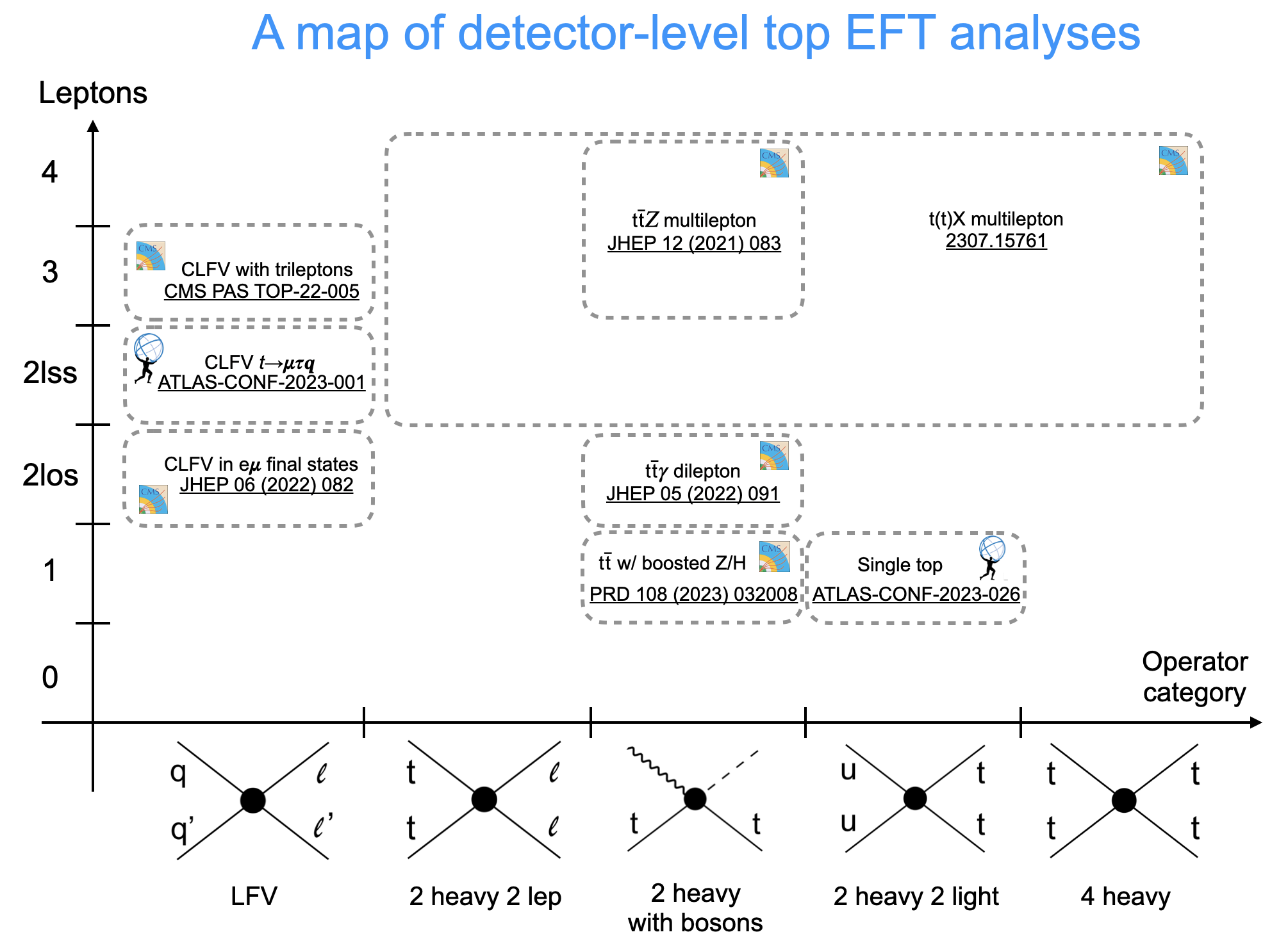}
\caption{Visual representation of some of the recent detector-level SMEFT analyses.}
\label{fig:map}
\end{figure}

Figure~\ref{fig:map} illustrates how these analyses map into the space of  recent detector-level SMEFT analyses.  The $x$-axis shows the categories of operators (as introduced in Section~\ref{intro}, though here the charged lepton flavor violating WCs are shown separately from the two-quark-two-lepton WCs), and the $y$-axis shows the final states studied in the analyses (categorized by the number of leptons included in the signal regions). 

\section{Summary}
Analyses in ATLAS and CMS utilize the SMEFT framework to search for the off-shell effects of heavy new physics. A variety of approaches are employed, ranging from indirect reinterpretations to direct approaches in which detector-level predictions are parameterized in terms of the SMEFT WCs. While these analyses have yet to observe any significant discrepancy with the SM, there remain many expansions and improvements that are being pursued; improvements in analysis techniques, improvements in EFT modeling,  the collection of more data, and combinations among analyses (both within the top quark sector and across multiple sectors) are leading to more comprehensive results that aim to improve the understanding of the ranges of WC values that are consistent with the observed data.


\bibliography{eprint}{}
\bibliographystyle{unsrt}
 
\end{document}